\documentclass[aps,10pt,final,
notitlepage, oneside, twocolumn, nobibnotes, nofootinbib,
superscriptaddress, showpacs, centertags, showkeys,
amssymb]{revtex4}

\usepackage{graphicx}
\usepackage{amsmath}

\begin{document}

\title{Sum rules for total cross-sections of hadron  photoproduction
on pseudoscalar mesons}

\author{S.~Dubni\v{c}ka \footnote{e-mail: fyzidubn@savba.sk}} \affiliation{Inst.\ of Physics,
Slovak Academy of\ Sciences, D\'ubravsk\'a cesta 9, 845 11
Bratislava, Slovak Republic}
\author{A.Z.~Dubni\v ckov\'a\footnote{e-mail: dubnickova@fmph.uniba.sk}}\affiliation{Dept.\ of Theor. Physics, Comenius University,\ Mlynsk\'a dolina,
842 48 Bratislava, Slovak Republic}
\author{E.~A.~Kuraev} \affiliation{Bogol'ubov\ Laboratory of Theoretical
Physics, JINR, Dubna, Russia}

%\date{\today}

\begin{abstract}
Sum rules  are derived relating mean squared charge radii of the
pseudoscalar mesons with the convergent integral of the difference
of hadron photoproduction cross-sections on pseudoscalar mesons.
 \vspace{1pc}
\end{abstract}

\pacs{11.55.Hx, 13.60.Hb, 25.20.Lj} \keywords{sum rule,
photoproduction, cross-section} \maketitle

\section{INTRODUCTION}

In the paper \cite{Bartos04}, considering the very high energy
electron-nucleon scattering with peripheral production of a
hadronic state $X$ moving closely to a direction of initial
nucleon and utilizing analytic properties of the forward Compton
scattering amplitudes on the nucleons, the new sum rule, relating
proton Dirac radius and anomalous magnetic moments of the proton
and the neutron to the convergent integral of the difference of
total proton and neutron photoproduction cross-sections, was
derived. In this paper we extend the previous method deriving sum
rules for various suitable couples of the members of the
pseudoscalar meson nonet, giving into a relation mean squared
charge radii with the convergent integral of the difference of
total hadron photoproduction cross-sections on the considered
pseudoscalar mesons.

\section{RELATION BETWEEN DIFFERENTIAL MESON ELECTROPRODUCTION AND
TOTAL HADRON PHOTOPRODUCTION CROSS-SECTIONS}

 Let us consider a very high energy peripheral electroproduction
process on pseudoscalar mesons $P$
\begin{equation}
e^-(p_1) + P(p) \to e^-(p_1') + X, \label{a1}
\end{equation}
where the produced pure hadronic state $X$ is moving closely to
the direction of the initial meson. Its matrix element in the one
photon exchange approximation takes the form
\begin{equation}
M = i \frac{\sqrt{4\pi \alpha}}{q^2} \bar u(p_1^{'})\gamma_\mu
u(p_1) <X \mid J_\nu \mid P>g^{\mu\nu}\label{a2}
\end{equation}
and $m^2_X = (p+q)^2$.

Now, by means of the method of equivalent photons \cite{Achie69},
examining the pseudoscalar meson in the rest, the electron energy
to be very high and the small photon momentum transfer, one can
express the differential cross-sections of the processes
(\ref{a1}) as a function of ${\bf q}^2$  through integral over the
total hadron photoproduction cross-sections on pseudoscalar
mesons.

Really, applying to (\ref{a2}) the Sudakov expansion \cite{Sud56}
of the photon transferred four-vector $q$
\begin{equation}\label{a3}
q=\beta_q \tilde{p}_1+\alpha_q\tilde{p}+q^{\bot} \quad
q_{\bot}=(0,0,{\bf q}),\quad q_{\bot}^2=-{\bf q}^2
\end{equation}
into the almost light-like vectors

\begin{equation}\label{a4}
\tilde{p}_1=p_1-m_e^2p/(2p_1p), \quad \tilde{p}=p-m_P^2p_1/(2p_1p),
\end{equation}
then using the Gribov prescription  \cite{Grib70} for the
numerator of the photon Green function

\begin{equation}
g_{\mu\nu}=g_{\mu\nu}^{\bot}+\frac{2}{s}(\tilde{p}_{\mu}\tilde{p}_{1\nu}+\tilde{p}_{\nu}\tilde{p}_{1\mu})
 \approx\frac{2}{s} \tilde{p}_\mu\tilde{p}_{1\nu}, \label{a5}
 \end{equation}
where $s=(p_1+p)^2\approx 2p_1p\gg Q^2 = -q^2$, as a consequence
of the electron energy in (\ref{a1}) to be very high and the
photon momentum transfer squared $t=q^2=-Q^2=-{\bf q^2}$ to be
small one, one obtaines for the corresponding cross-section
 \begin{eqnarray}\nonumber
&&d\sigma^{e^-P\to e^-X}= \frac{4\pi\alpha}{s(q^2)^2}
p_1^{\mu}p_1^{\nu}\sum_{X\neq P}\langle
P\mid J_\mu^{EM}\mid X\rangle^* \times \\
\label{a6} &\times& \langle X \mid J_\nu^{EM}\mid P\rangle d
\Gamma
\end{eqnarray}
with a summation through the created hadronic states $X$.

\begin{figure*}[hbt]
%\begin{center}
%\includegraphics[width=.45\columnwidth]{sum3.eps}
\includegraphics[scale=.7]{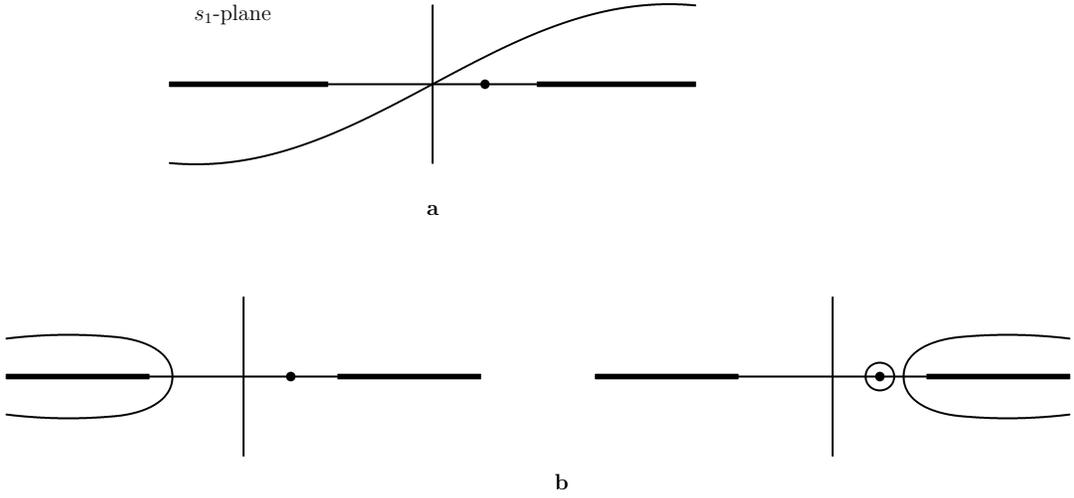}
\caption {\label{fig:1}Sum rule interpretation in $s_1$ plane.}
%\end{center}
\end{figure*}

Further, if the phase space volume of the final electron is
adjusted
\begin{equation}\label{a7}
\frac{1}{(2\pi)^3}\frac{d^3p'_1}{2\epsilon'_1}=\frac{1}{(2\pi)^3}d^4q
\delta[(p_1-q)^2]=\frac{1}{(2\pi)^3}\frac{d s_1}{2s}d^2q_\bot,
\end{equation}
in order to rewrite the final state phase-space volume in
(\ref{a6}) into the form
\begin{equation}\label{a8}
d\Gamma = \frac{d s_1}{2s(2\pi)^3}d^2{q_\bot}d\Gamma_X;
\end{equation}
with $$ s_1=2(qp)=m_X^2+{\bf q}^2-m^2_{P}=s\beta_q,$$ then the
current conservation condition ($\alpha_q\tilde p$ gives a
negligible contribution)
\begin{eqnarray}\label{a9}
&&q^\mu \langle X\mid J_\mu^{EM}\mid P\rangle\\
\nonumber&&\approx(\beta_q\tilde{p}_1+q_\bot)^\mu\langle X\mid
J_\mu^{EM}\mid P\rangle = 0,
\end{eqnarray}
is used in order to utilize in (\ref{a6}) the expression
\begin{eqnarray}\nonumber
&&\int p_1^\mu p_1^\nu \sum_{X\neq P}\langle P\mid J_\mu^{EM}\mid
X\rangle^*  \langle X \mid J_\nu^{EM}\mid P\rangle d \Gamma_X=\\
 \label{a10} &=& 2i\frac{s^2}{s_1^2}{\bf {q}}^2 Im \tilde {A}^{(P)}(s_1,{\bf
 {q}}),
\end{eqnarray}
with the amplitude $\tilde {A}^{(P)}(s_1,{\bf {q}})$ to be by a
construction only a part of the total forward virtual Compton
scattering amplitude ${A}^{(P)}(s_1,\bf {q})$ on pseudoscalar
mesons, which does not contain any crossing Feynman diagram
contributions, for a difference of corresponding differential
cross-sections of the electroproduction on $P$ and $P'$ (after
integration in (\ref{a6}) over $d\Gamma_X$, as well as over
$m^2_X$, i.e. over the variable $s_1$ to be interested only for
${\bf q}$ distribution) one finds

\begin{eqnarray} \label{a11}
&&\Big(\frac{d\sigma^{e^-P\to e^-X}}{d^2{\bf{q}}} -
\frac{d\sigma^{e^-P'\to
e^-X'}}{d^2{\bf{q}}}\Big)=\\
\nonumber &=&\frac{\alpha{\bf{q}}^2}
{4\pi^2}\int\limits_{s_1^{th}}^\infty\frac{d
s_1}{s_1^2[{\bf{q}}^2+(m_es_1/s)^2]^2}\times \\
\nonumber&\times& [Im \tilde{A^P}(s_1,{\bf{q}})- Im
\tilde{A^{P'}}(s_1,{\bf{q}})].
\end{eqnarray}

Finally, if one neglects the second term in square brackets of the
denominator of the integral in (\ref{a11})  (owing to the small
value of $m_e$ and high $s$ in comparison with $s_1$) and takes
the limit ${\bf{q^2}}\to 0 $ along with the expressions
$d^2\bf{q}=\pi d\bf{q^2}$ and $Im\tilde{A^{P}}(s_1,{\bf{q}})$=$
4s_1\sigma_{tot}^{\gamma{^*}P\to X}(s_1,{\bf{q}})$,  one comes to
the Weizs\"acker-Wiliams like relation
\begin{eqnarray} \label{a12}
&&{\bf{q}}^2\Big(\frac{d\sigma^{e^- P\to e^-X}}{d{\bf{q}}^2} -
\frac{d\sigma^{e^- P'\to e^-X'}}{d{\bf{q}}^2}\Big)_
{|_{{\bf{q}}^2\to 0}}=\\
\nonumber &=&\frac{\alpha}
{\pi}\int\limits_{s_1^{th}}^\infty\frac{d s_1}{s_1}
[\sigma_{tot}^{\gamma P\to X}(s_1)-\sigma_{tot}^{\gamma P'\to
X'}(s_1)]
\end{eqnarray}
between the difference of ${\bf q}^2$ - dependent differential
cross-sections of the processes (\ref{a1}) and the convergent
integral over the difference of the total  hadron photoproduction
cross-sections on pseudoscalar mesons.

\section{SUM RULES FOR PSEUDOSCALAR MESONS}

Now, let us investigate analytic properties of the forward Compton
scattering amplitude $\tilde{A} (s_1,\bf{q})$ in $s_1$-plane. They
consist in meson intermediate state pole at $s_1={\bf q^2}$, the
right-hand cut starting at the three meson threshold and the
$u_1$-channel left-hand cut. Defining the path integral $I$ (for
more detail see \cite{Kura81}) in $s_1$ plane
\begin{equation}\label{a13}
I=\int\limits_C d s_1 \frac{p_1^\mu p_1^\nu}{s^2}
\left(\tilde{A}^{(P)}_{\mu\nu}(s_1,{\bf{q}})-
\tilde{A}^{(P')}_{\mu\nu}(s_1,{\bf{q}})\right )
\end{equation}
from the gauge invariant light-cone projection
$p_1^{\mu}p_1^{\nu}\tilde{A}_{\mu\nu}$ of the part
$\tilde{A}_{\mu\nu}$ of the total Compton scattering tensor with
photon first absorbed and then emitted along the meson world line as
presented in Fig.1a and once closing the contour $C$ to upper
half-plane, another one to lower half-plane (see Figs. 1b), the
following sum rule
\begin{eqnarray}\nonumber
& &\pi (Res^{(P')}-Res^{(P)})
={\bf{q}}^2\int\limits_{r.h.}^\infty\frac{ds_1}
{s_1^2} [Im \tilde{A}^{(P)}(s_1,{\bf{q}})-\\
&-& Im \tilde{A}^{(P')}(s_1,{\bf{q}})] \label{a14}
\end{eqnarray}
appears with
\begin{equation}\label{a15}
Res^{(M)}=2\pi \alpha F^2_M(\bf{-q^2})
\end{equation}
to be the residuum of the meson intermediate state pole
contribution expressed through the pseudoscalar meson charge form
factor $F_M(\bf{-q^2})$ and the left-hand cut contributions
expressed by an integral of the difference $[Im
\tilde{A}^{(P)}(s_1,{\bf{q}}) - Im
\tilde{A}^{(P')}(s_1,{\bf{q}})]$ are mutually annulated. Then,
substituting (\ref{a15}) into (\ref{a14}) and  taking into account
(\ref{a11}) with $d^2{\bf q}=\pi d{\bf q}^2$, one comes to the
meson sum rules
\begin{eqnarray}
& &[F^2_{P'}({\bf{-q^2}})-F^2_{P'}(0)] -
[F^2_P({\bf{-q^2}})-F^2_P(0)] = \nonumber\\
&=&\frac{2}{\pi \alpha^2}({\bf{q^2}})^2\Big(\frac{d\sigma^{e^-
P\to e^-X}}{d{\bf{q}}^2} - \frac{d\sigma^{e^- P'\to
e^-X'}}{d{\bf{q}}^2}\Big)\label{a16},
\end{eqnarray}
where the left-hand side was renormalized in order to separate the
pure strong interactions from electromagnetic ones. Moreover,
substituting here for small values of ${\bf q}^2$ the relation
(\ref{a12}) and using the laboratory coordinate system by
$s_1=2m_P\omega$ and finally taking a derivative according to
${\bf{q}}^2$ of both sides for ${\bf q}^2=0$, one comes  to the
new sum rule relating meson mean squared charge radii to the
convergent integral of the difference of corresponding total
hadron photoproduction cross-sections on mesons

\begin{eqnarray}\label{a17}
&&\frac{1}{3}(F_P(0)\langle r_{P}^2 \rangle-F_{P'}(0)\langle r_{P'}^2 \rangle)=\\
\nonumber &=&\frac{2}{\pi^2\alpha}\int\limits_{{\omega_P}}^{\infty}
\frac{d\omega} {\omega}\big[\sigma_{tot}^{\gamma P\to X}(\omega)-
\sigma_{tot}^{\gamma P'\to X}(\omega)\big],
\end{eqnarray}
in which just a mutual cancelation of the rise of the latter cross
sections for $\omega \to \infty$ is achieved.

\section{APPLICATION TO VARIOUS COUPLES OF MESONS}

According to the SU(3) classification of existing hadrons there are
the following members of the ground state pseudoscalar meson nonet
$\pi^-$, $\pi^0$, $\pi^+$, $K^-$, $\bar K^0$, $K^0$, $K^+$, $\eta$,
$\eta'$. However, in consequence of CPT invariance the meson
electromagnetic form factors $F_P(-{\bf q^2})$ hold the following
relation
\begin{equation}
F_P({\bf -q^2})= - F_{\bar P}({\bf -q^2}),\label{a18}
\end{equation}
where $\bar P$ means antiparticle.

Since $\pi_0$, $\eta$ and $\eta'$ are true neutral particles,
their electromagnetic form factors are owing to the (\ref{a18})
zero in the whole region of a definition and therefore we exclude
them from further considerations.

If one considers couples of particle-antiparticle like
$\pi^{\pm}$, $K^{\pm}$ and $K^0$, $\bar K^0$, the left hand side
of (\ref{a16}) is owing to the relation (\ref{a18}) equal zero and
we exclude couples $\pi^{\pm}$, $K^{\pm}$ and $K^0$, $\bar K^0$
from further considerations as well.

If one  considers a couple of the isodoublet of kaons $K^+, K^0$
and $K^-, \bar K^0$, the following Cabibbo-Radicati \cite{Cab66}
like sum rules for kaons can be written
\begin{eqnarray}\label{a19}
\frac{1}{6}{\pi^2\alpha}\langle r_{K^+}^2\rangle
=\int_{\omega_{th}}^{\infty} \frac{d\omega}
{\omega}\left[\sigma_{tot}^{\gamma K^+\to X}(\omega)-
\sigma_{tot}^{\gamma K^0\to X}(\omega)\right]
\end{eqnarray}

\begin{eqnarray}\label{a20}
&&\frac{1}{6}{\pi^2\alpha}(-1)\langle r_{K^-}^2\rangle=\\\nonumber
&=&\int_{\omega_{th}}^{\infty} \frac{d\omega}
{\omega}\left[\sigma_{tot}^{\gamma K^-\to X}(\omega)-
\sigma_{tot}^{\gamma \bar K^0\to X}(\omega)\right]
\end{eqnarray}
in which the relation $\langle r_{K^+}^2\rangle=-\langle
r_{K^-}^2\rangle$ for kaon mean squared charge radii, following
directly from (\ref{a18}), holds and divergence of the integrals,
due to an increase of the total cross-sections
$\sigma_{tot}^{\gamma K^\pm\to X}(\omega)$ for large values of
$\omega$, is taken off by the increase of total cross-sections
$\sigma_{tot}^{\gamma K^0\to X}(\omega)$ and $\sigma_{tot}^{\gamma
\bar K^0\to X}(\omega)$, respectively.
   If besides the latter, also the relations
\begin{eqnarray}\label{a21}
\sigma_{tot}^{\gamma K^0\to X}(\omega)\equiv\sigma_{tot}^{\gamma
\bar K^0\to X}(\omega)\\\nonumber \sigma_{tot}^{\gamma K^+\to
X}(\omega)\equiv\sigma_{tot}^{\gamma K^-\to X}(\omega),
\end{eqnarray}
following from $C$ invariance of the electromagnetic interactions,
are taken into account, one can see the sum rule (\ref{a20}), as
well as all other possible sum rules obtained by combinations
$K^+\bar K^0, K^-K^0,$ to be contained already in (\ref{a19}).

The last possibility is a consideration of a couple of mesons
taken from the isomultiplet of pions and the isomultiplet of kaons
leading to the following less precise (in comparison with
(\ref{a19})) sum rules
\begin{eqnarray}\label{a22}
&&\frac{1}{6}{\pi^2\alpha}[(\pm1)\langle r_{\pi^{\pm}}^2\rangle-
(\pm1)\langle r_{K^{\pm}}^2\rangle]=\\
\nonumber &=&\int_{\omega_{th}}^{\infty} \frac{d\omega}
{\omega}\left[\sigma_{tot}^{\gamma \pi^{\pm}\to X}(\omega)-
\sigma_{tot}^{\gamma K^{\pm}\to X}(\omega)\right]
\end{eqnarray}

\begin{eqnarray}\label{a23}
 &&\frac{1}{6}{\pi^2\alpha}(\pm1)\langle r_{\pi^{\pm}}^2\rangle =\\
\nonumber &=&\int_{\omega_{th}}^{\infty} \frac{d\omega}
{\omega}\left[\sigma_{tot}^{\gamma \pi^{\pm}\to X}(\omega)-
\sigma_{tot}^{\gamma K^0\to X}(\omega)\right],
\end{eqnarray}
as there is no complete annulation of the left-hand cut
contributions in (\ref{a14}) due to a larger difference in the
masses of joining pairs of particles.

   The latter assertion can be roughly confirmed as follows. The
left-hand cut contribution in (\ref{a14}) has no direct
interpretation in terms of a cross-section. Nevertheless, it can
be associated (for more detail see ref. \cite{Kura06}) with
contribution to the cross-sections of $M \bar M$ meson pair
electroproduction on considered target meson $M$, arising from
taking into account the identity of final state mesons. Then the
left-hand cut contribution to the derivative according to $\vec
q^2$ at $\vec q^2 = 0$ of scattering amplitudes entering sum rules
have an order of magnitude

\begin{equation}
I= \frac{g^4}{(2\pi)^3 s_{1 max}}, \quad s_{1
max}=max[m_\rho^2,8m_M^2]
\end{equation}
where $g$ is the strong coupling constant of the $\rho$-meson to
the considered meson and $m_M$ is the mass of the target meson.
Taking the PDG \cite{Eid04} typical value for the total
cross-section of scattering of a pion on proton to be
$\sigma_{tot}^{\pi N} \equiv 20 \quad [mb]$, then $\rho$-meson
$t$-channel contribution and $s_{1 max}=m_\rho^2$ in the case of
the target meson is charged pion and $s_{1 max}=8m_K^2$ in the
case of the target meson is kaon, we have  $I_\pi \approx 0.081
\quad [mb]$ and $I_K \approx 0.024 \quad [mb]$ respectively, which
confirm above mentioned statement.

   Now taking the experimental values \cite{Eid04}\\ $(\pm1)\langle
r_{\pi^{\pm}}^2\rangle=+0.4516\pm0.0108 \quad [fm^2]$\\
and\\
$(\pm1)\langle r_{K^{\pm}}^2\rangle=+0.3136\pm0.0347 \quad [fm^2]$\\
one comes to the conclusion that in average
\begin{eqnarray}
[\sigma_{tot}^{\gamma \pi^{\pm}\to X}(\omega)-
\sigma_{tot}^{\gamma K^{\pm}\to X}(\omega)] > 0\\ \nonumber
[\sigma_{tot}^{\gamma K^-\to X}(\omega)- \sigma_{tot}^{\gamma \bar
K^0\to X}(\omega)] > 0
\end{eqnarray}
from where the following inequalities for finite values of
$\omega$ in average follow
\begin{eqnarray}
\sigma_{tot}^{\gamma \pi^{\pm}\to X}(\omega)> \sigma_{tot}^{\gamma
K^{\pm}\to X}(\omega) > \sigma_{tot}^{\gamma \bar K^0\to
X}(\omega) > 0.
\end{eqnarray}

Subtracting up (\ref{a19}) or (\ref{a20}) from the relation
(\ref{a23}) the sum rule (\ref{a22}) is obtained, what
demonstrates a mutual consistency of all considered sum rules.
They have been derived in analogy with a derivation
\cite{Bartos04} of the sum rule for a difference of proton and
neutron total photoproduction cross-sections, which are fulfilled
with a very high precision. Therefore we believe that also the sum
rules for total cross-sections of hadron photoproduction on
pseudoscalar mesons presented in this paper are correct. However,
the final word is always given by experimental tests.

   The experimental test of the derived sum rules can be practically carried out
if there are known the total hadron photoproduction cross-sections
on pions and kaons as a function of energy, which, however, are
missing till now. Nevertheless, the idea of a conversion of the
electron beams of linear $e^+e^-$ colliders into photon beams,
using the process of the backward Compton scattering of laser
light off the high energy electrons, which is known \cite{Gin81}
already for few decades, together with a real possibility of a
production of enough intensive beams of pions \cite{Fra06} provide
a real chance for measurements of the total hadron photoproduction
cross-sections on charged pions and kaons, and as a result also
the experimental test of the sum rules derived in this paper.

\section{CONCLUSIONS}

Considering the very high energy  peripheral electron pseudoscalar
meson scattering with a production of a hadronic state $X$ moving
closely to the direction of initial meson, then utilizing analytic
properties of the forward Compton scattering amplitude on the same
meson, for the case of small transferred momenta new
Cabibbo-Radicati \cite{Cab66} like sum rules,  relating the
corresponding meson mean squared charge radii with the convergent
integral over a difference of the total hadron photoproduction
cross-sections on mesons are derived. Unlike the sum rules
(\ref{a19}), (\ref{a20}), (\ref{a22}) and (\ref{a23}), derived in
this paper, there could be difficulties with experimental
verification of the Cabibbo-Radicati sum rule due to appearance of
the sum of total photoproduction cross-sections on pions with the
transition always to a hadronic state with a specific isospin
state and moreover, the convergence of the corresponding integral
is questionable.

The work was partly supported by Slovak Grant Agency for Sciences,
Grant 2/4099/25  (S.D. and A.Z.D.). The authors (S.D. and A.Z.D)
would like to thank to TH Division of CERN  for a warm hospitality
where the paper was finished.

\end{document}